\documentclass{article}
\usepackage[margin=0.75in]{geometry}
\usepackage{graphicx}
\usepackage{hyperref}
\usepackage[nosectionbib]{apacite}
\usepackage{verbatim}

\title{Exposure to Social Engagement Metrics \\ Increases Vulnerability to Misinformation}

\author{Mihai Avram$^{1,3}$, Nicholas Micallef$^{2}$, Sameer Patil$^{3}$, Filippo Menczer$^{1,3}$ \\ $^{1}$Observatory on Social Media, Indiana University, Bloomington, IN, USA \\ $^{2}$Center for Cybersecurity, New York University, Abu Dhabi, United Arab Emirates\\ $^{3}$Luddy School of Informatics, Computing, and Engineering, Indiana University Bloomington, IN, USA}

\date{}

\begin{document}

\maketitle

Topics: Misinformation; social media; social engagement; news literacy; game  

\section*{Article's lead}

News feeds in virtually all social media platforms include engagement metrics, such as the number of times each post is liked and shared. We find that exposure to these social engagement signals increases the vulnerability of users to low-credibility information. This finding has important implications for the design of social media interactions in the misinformation age. To reduce the spread of misinformation, we call for technology platforms to rethink the display of social engagement metrics. Further research is needed to investigate whether and how engagement metrics can be presented without amplifying the spread of low-credibility information.


\section*{Research questions}
\begin{itemize}
\item What is the effect of exposure to social engagement metrics on people's propensity to share content?
\item Does exposure to high engagement metrics increase the chances that people will like and share questionable content and/or make it less likely that people will engage in fact checking of low-credibility sources?
\end{itemize}

\section*{Essay summary}     

\begin{figure}
\centering
  \includegraphics[width=0.75\textwidth]{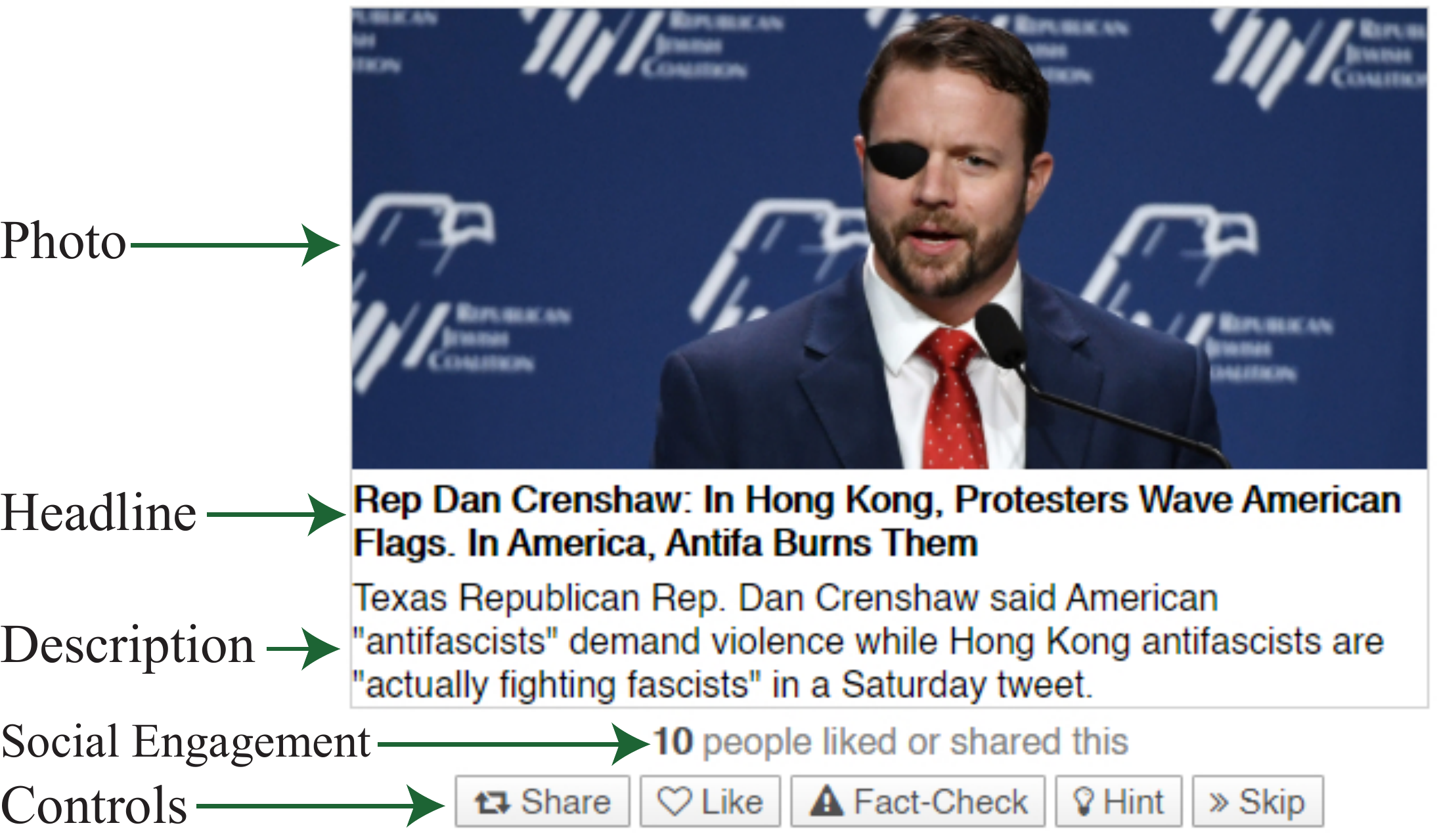}
  \caption{A news post in the social media feed simulated by the game.}
\label{fig:feeddetails}
\end{figure}

\begin{itemize}

    \item We investigated the effect of social engagement metrics on the spread of information from low-credibility sources using Fakey~\footnote{\url{https://fakey.iuni.iu.edu/}}, a news literacy game that simulates a social media feed (Figure~\ref{fig:feeddetails}). The game presents users with actual current news articles from mainstream and low-credibility media sources. A randomly generated social engagement metric is displayed with each presented article. Users are instructed to share, like, fact check, or skip articles. 
    
    \item From a 19-month deployment of the game, we analyzed game sessions by over 8,500 unique users, mostly from the US, involving approximately 120,000 articles, half from low-credibility sources. 
    
    \item Our findings show that displayed engagement metrics can strongly influence interaction with low-credibility information. The higher the shown engagement, the more prone people were to share questionable content and less to fact check it.
    
    \item These findings imply that social media platforms must rethink whether and how engagement metrics should be displayed such that they do not facilitate the spread of misinformation or hinder the spread of legitimate information. Further research is needed to guard against malicious tampering with engagement metrics at an early stage and to design educational interventions that teach users to prioritize trustworthiness of news sources over social engagement metrics.

\end{itemize}


\section*{Implications}

Online misinformation is a critical societal threat in the current digital age, and social media platforms are a major vehicle used to spread it~\cite{guess2019less,lazer2018thescience,hameleers2020picture}. As an illustration, the International Fact-Checking Network found more than 3,500 false claims related to the coronavirus in less than 3.5 months.\footnote{\url{https://poynter.org/coronavirusfactsalliance}} Viral misinformation can cause serious societal harm in multiple ways: affecting public health~\cite{sharma2020coronavirus}, influencing public policy~\cite{lazer2018thescience}, instigating violence~\cite{arif2018acting,starbird2014rumors}, spreading conspiracies~\cite{samory2018conspiracies}, reducing overall trust in authorities~\cite{gupta2014tweetcred,shin2017partisan,vosoughi2018the}, and increasing polarization and conflict~\cite{stewart2018examining}.

The growing societal impact of misinformation has driven research on technical solutions to detect and in some cases --- depending on platform policies --- stop actors that generate and spread such content. The techniques have leveraged network analytics~\cite{Truthy_icwsm2011class,jin2013epidemiological}, supervised models of automated behavior~\cite{socialbots-CACM,botornot_icwsm17,yang2019arming,Yang2020botometer-lite,hui2019botslayer}, time series analysis to detect promoted campaigns~\cite{campaigns2017}, and natural language processing for flagging factually incorrect content~\cite{prezrosas2017automatic,kumar2016disinformation}. On the user interface side, researchers have explored the use of credibility indicators to flag misinformation and alert users~\cite{clayton2019real}. Such credibility indicators can lead to a reduction in sharing the flagged content~\cite{yaqub2020effects,pennycook2019implied,pennycook2020fighting,nyhan2019taking}.

Studies have explored the role of environmental, emotional, and individual factors in online contagion~\cite{kramer2014experimental, ferrara2015measuring,coviello2014detecting,grinberg2019fake,yaqub2020effects}. 
However, there has been little empirical research on the effects of current elements of social media feeds on the spread of misinformation~\cite{hameleers2020picture,shen2019fake}. To address this gap, we empirically investigated how the spread of low-credibility content is affected by exposure to typical social engagement metrics, i.e., the numbers of likes and shares shown for a news article. We found near-perfect correlations between displayed social engagement metrics and user actions related to information from low-credibility sources. 
We interpret these results as showing that social engagement metrics are doubly problematic for the spread of low-credibility content: high levels of engagement make it less likely that people will scrutinize potential misinformation, at the same time making it more likely that they will like or share it. 
For example, we recently witnessed a campaign to make the ``Plandemic'' disinformation video go viral. Our results tell us that people are more likely to endorse the video without bothering to verify its content, simply because they see that many other people shared it. 
In other words, exposure to engagement metrics in social media amplifies our vulnerability to questionable content.

To interpret these findings, consider that the probability of sharing a piece of information grows with the number of times one is exposed to it, a phenomenon called \emph{complex contagion}~\cite{romero2011differences,monsted2017evidence}. Engagement metrics are proxies for multiple exposures, therefore they are intended to provide signals about the importance, relevance, and reliability of information --- all of which contribute to people's decisions to consume and share the information. In other words, being presented with high engagement metrics for an article mimics being exposed to the article multiple times: the brain is likely to assess that the article must be worthy of attention because many independent sources have validated the news article by liking or sharing it. 

A key weakness in the cognitive processing of engagement metrics is the assumption of independence; an entity can trick people by maliciously boosting engagement metrics to create the \emph{perception} that many users interacted with an article. In fact, most disinformation campaigns rely on inauthentic social media accounts to tamper with engagement metrics, creating an initial appearance of virality that becomes reality once enough humans are deceived~\cite{shao2018spread}. To prevent misinformation amplified by fake accounts from going viral, we need sophisticated algorithms capable of early-stage detection of coordinated behaviors that tamper with social engagement metrics~\cite{hui2019botslayer,Yang2020botometer-lite,pacheco2020uncovering}. 

Our findings hold important implications for the design of social media platforms. Further research is needed to investigate how alternative designs of social engagement metrics could reduce their effect on misinformation sharing (e.g., by hiding or making engagement less visible for certain posts), without negatively impacting the sharing of legitimate and reliable content. A good trade-off between these two conflicting needs will require a systematic investigation of news properties that can help determine differential display of engagement metrics. Such properties may include the type of sources (e.g., whether claims originate from unknown/distrusted accounts or low-credibility sources) and the type of topics (e.g., highly sensitive or polarizing topics with a significant impact on society). 

Further research is also needed to design literacy campaigns (such as Fakey~\footnote{\url{https://fakey.iuni.iu.edu/}}, a news literacy game that simulates a social media feed) that teach users to prioritize trustworthiness of sources over engagement signals when consuming content on social media. Studies could investigate the possibility of introducing intermediary pauses when consuming news through a social media feed~\cite{fazio2020pausing} and limiting automated or high-speed sharing. A comprehensive literacy approach to reduce the vulnerability of social media users to misinformation may require a combination of these interventions with others, such as inoculation theory~\cite{roozenbeek2020prebunking,roozenbeek2019fake,roozenbeek2019the, basol2020good}, civic online reasoning~\cite{mcgrew2020learning}, critical thinking~\cite{lutzke2019priming}, and evaluation of news feeds~\cite{nygren2019diversity}.


\section*{Findings}

\subsection*{Finding 1: High levels of social engagement results in lower fact checking and higher liking/sharing, especially for low-credibility content.}  

For each article shown in the game, the user is presented a photo, headline, description, and a randomly generated social engagement level. Based on this information, the user can share, like, or fact check the article (Figure~\ref{fig:feeddetails}). To earn points in the game, the user must share or like articles from mainstream sources and/or fact check articles from low-credibility sources. The social engagement level shown with the article provides an opportunity to investigate its effect on behaviors that result in the spread of questionable content.

We measured the correlation between the social engagement metric $\eta$ displayed to users and the rates at which the corresponding articles from low-credibility sources were liked/shared or fact checked by the users. Given the realistically skewed distribution of $\eta$ values, we sorted the data into logarithmic bins based on the shown social engagement levels. For each bin $\lfloor \log_{10}(\eta + 1) \rfloor$, we calculated the liking/sharing and fact-checking rates across articles and users. We measured correlation using the non-parametric Spearman test as the data is not normally distributed. We found a significant positive correlation between social engagement level and liking/sharing (Spearman $\rho=0.97$, $p<0.001$) and a significant negative correlation between social engagement level and fact checking (Spearman $\rho=-0.97$, $p<0.001$) for articles from low-credibility sources. 

We found similar relationships between social engagement levels and user behaviors for mainstream news article as well, however the correlations are less strong: $\rho=0.66$ for liking/sharing and $\rho=-0.62$ for fact checking. 

\subsection*{Finding 2: People are more vulnerable to low-credibility content with high social engagement.}

\begin{figure}
\centering
  \includegraphics[width=0.75\textwidth]{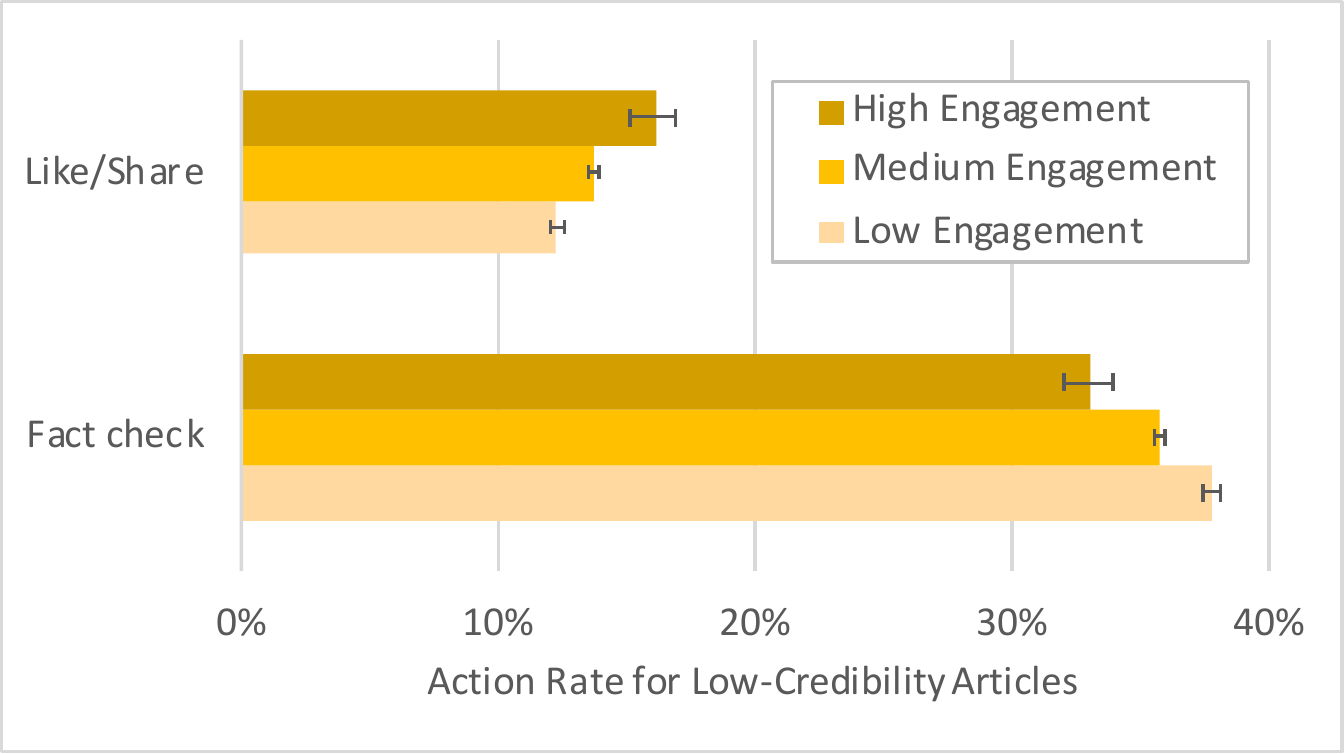}
  \caption{Mean rates of liking/sharing and fact checking low-credibility articles, categorized by social engagement level (see text). Error bars represent the standard error rate.}
\label{fig:userengagement}
\end{figure}

The previous finding is at the population level, aggregating across users. To delve further into the effect of social engagement exposure on individual users, we analyzed whether different social engagement levels influenced each user's liking/sharing and fact-checking rates for articles from low-credibility sources. For this analysis, we treated each user as an independent entity and categorized engagement into three levels: low ($0 \leq \eta < 10^2$), medium ($10^2 \leq \eta < 10^5$), and high ($10^5 \leq \eta \leq 10^6$). 
For each user, we counted the number of low-credibility articles to which they were exposed within each social engagement bin. We then calculated the corresponding proportions of these articles that each user liked/shared or fact checked.
Figure~\ref{fig:userengagement} plots the mean liking/sharing and fact-checking rates for low-credibility articles. Although users were more likely to fact check than like or share low-credibility content, Figure~\ref{fig:userengagement} shows that the trends observed at the population level held at the individual level as well.

Since the data is not normally distributed ($p<0.05$ using the Shapiro-Wilk test for normality), we used the Kruskal-Wallis test to compare differences between the three bins of social engagement levels. The test revealed a statistically significant effect of social engagement levels: fact checking ($\chi^2(2) = 214.26$, $p<0.001$) and liking/sharing ($\chi^2(2) = 417.14$, $p<0.001$) rates for low-credibility articles differed across the bins. To determine which levels of social engagement impacted the rates at which low-credibility articles were liked/shared or fact checked, we conducted post-hoc Mann-Whitney tests with Bonferroni correction for multiple testing across all pairs of social engagement bins and found that liking/sharing as well as fact-checking rates were statistically significantly different across all pairings ($p<0.001$).

We employed the same approach to examine liking/sharing and fact-checking rates for mainstream articles across the three bins of social engagement levels. Similar to low-credibility articles, the Kruskal-Wallis test revealed a statistically significant effect of social engagement level on liking/sharing ($\chi^2(2) = 161.80$, $p<0.001$) and fact checking ($\chi^2(2) = 576.37$, $p<0.001$) rates for mainstream articles.


\section*{Methods}
 
\subsection*{Social media simulation}

To conduct our experiment investigating the effect of exposure to social engagement metrics on susceptibility to questionable content, we developed and deployed Fakey~\footnote{\url{https://fakey.iuni.iu.edu/}}, an online news literacy game that simulates fact checking on a social media feed. The user interface of the game mimics the appearance of Facebook or Twitter feeds for players who log into the game through those platforms. The game provides users with batches of ten news articles in the form of a news feed, as shown in Figure~\ref{fig:feeddetails}. Each article consists of elements that are typically displayed by popular social media platforms: photo, headline, description, and social engagement metrics. 

For each article from mainstream as well as low-credibility sources, the game displays a single social engagement metric about the combined number of shares and likes. Not having separate metrics for shares and likes decreases the cognitive workload for game players and simplifies analysis. Engagement values are randomly drawn from an approximately log-normal distribution with a maximum possible value (cutoff) of $\eta=10^6$. The distribution is such that roughly 69\% of the articles would display engagement values $\eta > 10^2$ and roughly 3\% would display values $\eta > 10^5$. Although the simulated engagement in the game is not drawn from empirical data, the metric numbers shown have a heavy tail similar to those typically observed in social media~\cite{vosoughi2018the}.   

Below each article is a set of action buttons to share, like, fact check, or skip the article or use a hint. Before playing the game, users are instructed that clicking \emph{Share} is equivalent to endorsing an article and sharing it with the world, clicking \emph{Like} is equivalent to endorsing the article, and clicking \emph{Fact Check} signals that the article is not trusted. After playing one round of ten articles, users have the option to play another round or check a leader-board to compare their skill with other players.

\subsection*{Content selection}

We follow the practice of analyzing content credibility at the domain (website) level rather than the article level~\cite{lazer2018thescience,shao2018spread,Shao2018anatomy,grinberg2019fake,pennycook2019fighting,bovet2019influence}. 
Each article in the game is selected from one of two types of news sources: mainstream and low-credibility. 

For mainstream news, we manually selected 32 sources with a balance of moderate liberal, centrist, and moderate conservative views: \textit{ABC News Australia, Al Jazeera English, Ars Technica, Associated Press, BBC News, Bloomberg, Business Insider, Buzzfeed, CNBC, CNN, Engadget, Financial Times, Fortune, Independent, Mashable, National Geographic, New Scientist, Newsweek, New York Magazine, Recode, Reuters, Techcrunch, The Economist, The Guardian, The New York Times, Next Web, Telegraph, Verge, The Wall Street Journal, The Washington Post, Time, USA Today.} Current articles are provided by the News API.\footnote{\url{https://newsapi.org}}

The set of low-credibility sources was selected based on flagging by various reputable news and fact-checking organizations~\cite{shao2018spread,Shao2018anatomy}. These sources tend to publish fake news, conspiracy theories, clickbait, rumors, junk science, and other types of misinformation. The articles are provided by the Hoaxy API.\footnote{\url{http://rapidapi.com/truthy/api/hoaxy}}

For each round, the game randomly selects five articles each from mainstream and low-credibility sources. For any given source, any article returned by the News or Hoaxy API is shown to the user regardless of topic, without further selection or filtering except for ensuring that the same story is not shown to the same player multiple times across rounds. 

\subsection*{Data collection}

The game is available online through a standard web interface and as a mobile app via the Google Play Store and the Apple App Store. The mobile app is available in English-speaking countries: United States, Canada, United Kingdom, and Australia. People from other countries can still play the game through the web interface. 

The present analysis is based on data from a 19-month deployment of the game, between May 2018 and November 2019. During this period, we advertised the game through several channels, including social media (Twitter and Facebook), press releases, conferences, keynote presentations, and word of mouth. We recorded game sessions involving approximately 8,606 unique users\footnote{We used analytics to aggregate anonymous sessions by the same person. However, this approach cannot ascribe anonymous sessions to a single person with complete certainty. Therefore, we cannot provide a precise number of unique users.} and 120,000 news articles, approximately half of which from low-credibility sources. We did not collect demographic information, but we collected anonymous data from Google Analytics embedded within the game's hosting service. Participants originated from the United States (78\%), Australia (8\%), UK (4\%), Canada (3\%), Germany (3\%), and Bulgaria (2\%).

\subsection*{Limitations}

Our news literacy app emulates relevant interface elements of popular social media platforms such as Facebook and Twitter, without the ethical concerns of real-world content manipulation~\cite{kramer2014experimental}. Yet, conducting the study in a simulated game environment rather than an actual platform presents clear limitations --- the experience and context are not identical. For example, to limit the cognitive burden on players, we capture only Like and Share actions; these were the earliest ones deployed on social media platforms and as such are most common across platform and most familiar to users. 

The even mix of articles from mainstream and low-credibility sources is not necessarily representative of the proportion of misinformation to which social media users are exposed in the wild. 
The fact-checking game also primes users to \emph{expect} misinformation, potentially making it more likely to be spotted.
These factors might make users more suspicious within the game compared to the real world, increasing fact-checking rates. However, there is no reason to believe that these effects impact results about engagement signals.

While this study is focused on user interaction elements, other factors related to users and content can affect the spread of questionable content. To respect player privacy, we chose not to collect any user information apart from game analytics. However, knowledge about the background of the users (e.g., education, demographics, political affiliation) might provide further insight into vulnerability to misinformation. Similar refinements in insight would be provided by examining which types of content are more likely to be influenced by engagement metrics. These are important avenues for future research.


\section*{Acknowledgments}

We are grateful to Alessandro Flammini for inspiring the social engagement exposure experiment and Chengcheng Shao and Clayton A. Davis for technical support during platform development.

\section*{Bibliography}

\bibliographystyle{apacite}
\renewcommand{\refname}{}
\vspace{-1.2cm}
\bibliography{main.bib}

\section*{Funding}

M.A. and F.M. were supported in part by the Democracy Fund, Craig Newmark Philanthropies, and Knight Foundation. 
The funders had no role in study design, data collection and analysis, decision to publish, or preparation of the manuscript.

\section*{Competing interests}

The authors have no competing interests to declare.

\section*{Ethics}

The mechanisms and procedures reported in this article were reviewed and approved by the Institutional Review Board (IRB) of the authors' institution.  

\end{document}